\documentclass[10pt, journal]{IEEEtran}
\usepackage{algorithmicx}
\usepackage[ruled,vlined,linesnumbered]{algorithm2e}
\usepackage{hhline}
\usepackage{amsmath,mathtools}
\usepackage{amsfonts,amssymb}
\usepackage{mathrsfs}
\usepackage{caption}
\usepackage{enumitem}
\usepackage{multirow}
\usepackage{graphicx} 
\usepackage{enumitem,color}
\usepackage{algpseudocode}

\begin{document}
%

\title{Deep Reinforcement Learning for End-to-End Network Slicing: Challenges and Solutions}

\author{
        Qiang Liu,
        Nakjung Choi,
        Tao Han\vspace{-0.3in}
\IEEEcompsocitemizethanks{
        \IEEEcompsocthanksitem Qiang Liu is with the School of Computing, University of Nebraska-Lincoln.
        E-mail: qiang.liu@unl.edu
        \IEEEcompsocthanksitem Nakjung Choi is with Nokia Bell Labs.
        E-mail: nakjung.choi@nokia-bell-labs.com
        \IEEEcompsocthanksitem Tao Han is with the Department of Electrical and Computer Engineering, New Jersey Institute of Technology.
        E-mail: tao.han@njit.edu
}
}

\maketitle

\begin{abstract}
5G and beyond is expected to enable various emerging use cases with diverse performance requirements from vertical industries. \textcolor{black}{To serve these use cases cost-effectively, network slicing plays a key role in dynamically creating virtual end-to-end networks according to specific resource demands}. A network slice may have hundreds of configurable parameters over multiple technical domains that define the performance of the network slice, which makes it impossible to use traditional model-based solutions to orchestrate resources for network slices. In this article, we discuss how to design and deploy deep reinforcement learning (DRL), a model-free approach, to address the network slicing problem. First, we analyze the network slicing problem and present a standard-compliant system architecture that enables DRL-based solutions in 5G and beyond networks. Second, we provide an in-depth analysis of the challenges in designing and deploying DRL in network slicing systems. Third, we explore multiple promising techniques, i.e., safety and distributed DRL, and imitation learning, for automating end-to-end network slicing. 
\end{abstract}

\section{Introduction}
\label{sec:introduction}
5G and beyond will be a catalyst to digitalize the economy and contribute toward global digital transformation. The explosion of networking connections and mobile data will dramatically increase the complexity of the network. An increasing number of new use cases will be enabled for various industries such as automotive, manufacturing, logistics, and energy~\cite{5GSlicing}. These new use cases have highly diverse, and even conflicting, communication and networking requirements such as latency, data rates, availability, and reliability. The growing network complexity and service diversity challenge network operators to dynamically orchestrate and coordinate network resources to offer a different mix of capacities for supporting services with diverse requirements simultaneously.

Since it is not economically viable to build a dedicated network for each type of service, network slicing emerges as a key technology in 5G wireless networks for efficiently supporting highly diverse use cases~\cite{5GSlicing}. \textcolor{black}{Network slicing allows network operators to run multiple logical networks (aka. network slices) on top of common physical network infrastructures. For each network slice, customized functionalities and network operations can be implemented according to the specific need of an individual use case, e.g., enhanced mobile broadband (eMBB), ultra reliable low latency communications (URLLC) and massive machine type communication (mMTC)}. For example, a network slice can be customized to support IoT services for a large number of devices operated at low data rates. At the same time, another network slice can be tailored to provision latency-critical services such as vehicle-to-vehicle communications and smart grid controls. 

\begin{figure}[!t]
	\centering	
	\includegraphics[width=3.0in]{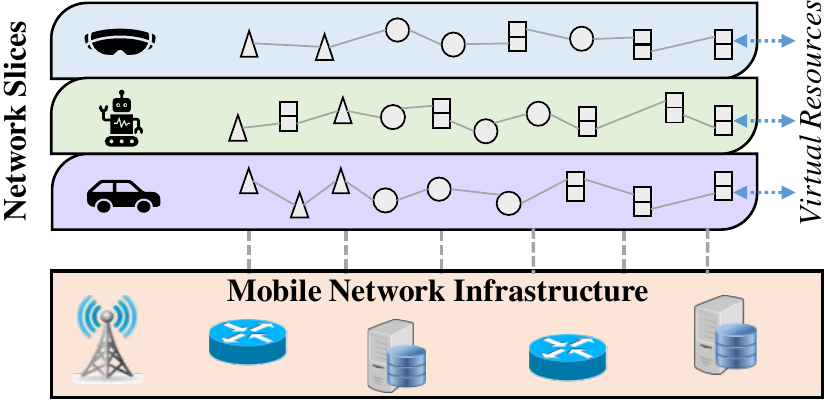}
	\caption{\small \textcolor{black}{An illustration of end-to-end network slicing.}}
	\label{fig:network_slicing}
\end{figure}

\textcolor{black}{Network slices usually span across multiple technical domains of the network, as shown in Fig.~\ref{fig:network_slicing}, e.g., radio access networks, transportation networks, core networks, and edge and cloud computing. Hence, end-to-end resource orchestration is essential in network slicing to dynamically manage resource allocations to different slices in multiple domains for achieving optimized performances}. A network slice may have hundreds of configuration parameters defining its functions and performance according to the service level agreement (SLA). These configurations are connected to the underlying resource demands in different domains. Thus, it is impractical to develop a closed-form model of the network performance versus the joint resource allocations in multiple domains.

\textcolor{black}{
Recent advances in ML, especially deep learning (DL) and deep reinforcement learning (DRL), have demonstrated great potential to learn and understand complex and high-dimensional correlations by leveraging artificial neural network (ANN) architectures. 
ML as a model-free approach requires no prior knowledge in advance, which can automatically learn the complex and unknown correlations in network slicing.
Thus, there is increasing popularity of exploring ML to automate network slicing~\cite{khan2020network, hurtado2022deep, kafle2019automation}, e.g., admission control and resource management.
Recent works reveal that the resource orchestration in network slicing has Markov property~\cite{liu2020edgeslice}, i.e., the orchestration decision at the current time influences not only the current performance of slices but also the future network states, e.g., queuing.
As a result, the resource orchestration turns out to be a Markov decision process (MDP), which cannot be resolved with DL techniques.
Hence, DRL becomes the most suitable approach to deal with the MDP and intelligently manage the resource orchestration in network slicing.
}

To facilitate ML-based network management solutions, the open-radio access network (O-RAN) alliance proposes a general framework of ML procedure which incorporates ML components, e.g., supervised and unsupervised learning, within multiple network functions~\cite{o-ran}.
ETSI introduces experiential networked intelligence (ENI) to enable context-aware artificial intelligence based on the “observe-orient-decide-act” control model. It supports adaptive and intelligent service operation and management for network operators by integrating network function virtualization (NFV) and software-defined network (SDN) controllers.
3GPP suggests the management loop of "observation-analytics-decision-execution" by leveraging the management data analytics (MDA), which provides the capability of processing and analyzing the raw data related to network and service events and status, e.g., performance measurements, and QoE reports.

Following these initiatives, this article dives deep into DRL-based network slicing solutions. First, we introduce the end-to-end resource orchestration problem in network slicing and show that the problem can be naturally formulated as a Markov decision process (MDP). To deploy DRL-based network slicing, a new end-to-end network system architecture is engineered with the network layer, orchestration layer and intelligence layer. \textcolor{black}{Second, we discuss the challenges of designing DRL-based network slicing solutions from the perspectives of both algorithms and system. Third, we explore and envision multiple promising techniques, i.e., safety and distributed DRL, and transfer learning, to address these challenges in automating the end-to-end network slicing. }

\section{DRL for Network Slicing}
\label{sec:architecture}

\textcolor{black}{A fundamental research problem in network slicing is the design of end-to-end resource orchestration that jointly orchestrates resources in multiple domains, e.g., radio access networks and edge computing, based on service requirements of network slices.} In this section, we analyze the end-to-end resource orchestration problem and present the system architecture for DRL-based network slicing. 

\subsection{Resource Orchestration Problem}
The rationale behind using DRL is that the end-to-end resource orchestration problem can be reasonably formulated as a Markov decision process (MDP) when considering the temporal dependency of the resource orchestration and network performance~\cite{liu2020edgeslice}. \textcolor{black}{A MDP can be denoted by $\langle S, A, r, P, \mu\rangle$, where $\textbf{s}^t \in S$ is defined as the \textit{\textbf{state}} that describes the network status and traffic load distributions in the current time slot $t$}. $\textbf{a}^t \in A$ is defined as \textit{\textbf{action}} that describes how resources are orchestrated on every edge node, e.g., radio base stations and edge servers, for every network slices in time slot $t$. $r: S\times A \to \mathbb{R}$ is defined as the \textit{\textbf{reward function}} that models the performance of the network system (e.g., resource usage).
\textcolor{black}{$c: S\times A \to \mathbb{C}$ is defined as the \textit{\textbf{cost function}} that models the constraints of the network system (e.g., SLA requirement).}
$P: S\times A\times S\to \mathbb{R}$ is defined as the \textit{\textbf{state transition function}} describes the underlying temporal dynamics of the state depending on the current action and \textit{\textbf{initial state distribution}} $\mu: S\to [0,1]$. The resource orchestration action is denoted by $\textbf{a}^t = \{a^t_{m,k}\}$ where $a^t_{m,k}$ is the amount of type-$m$ resource allocated to slice $k$ in time slot $t$. 
A resource orchestration policy $\pi: S \to Pr(A)$ prescribes a probability distribution over actions given the current state. \textcolor{black}{Given policy $\pi$, the end-to-end networking system receives a sequence of rewards and costs (i.e., network resource usage and system performance) $r^1, r^2, ...$ and $c^1, c^2, ...$. The end-to-end orchestration problem is to find the policy that maximizes the long-term system reward, i.e., $\arg\max_\pi \mathbb{E}^{\pi_\theta} [\sum_{t=0}^\infty \gamma^t r^t(\textbf{s}^t, \textbf{a}^t), \; s.t. \; c^t(\textbf{s}^t, \textbf{a}^t)\leq 0]$ where $\gamma$ is the discount factor that balances immediate and future rewards.}

\begin{figure*}[!t]
	\centering	
	\includegraphics[width=6.5in]{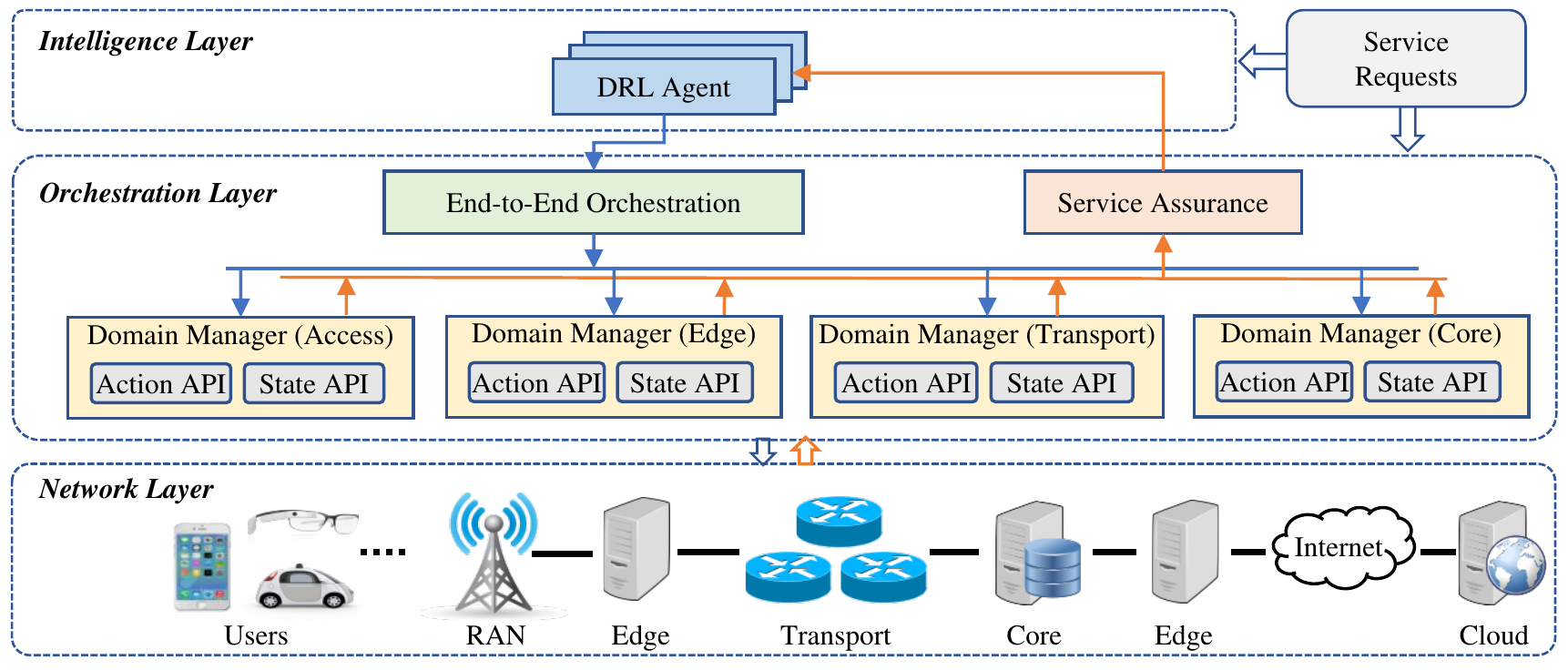}
	\caption{\small DRL-powered end-to-end network slicing system architecture.}
	\label{fig:network_architecture}
\end{figure*}

\subsection{DRL-based Network Slicing System Architecture}

\textcolor{black}{To facilitate DRL-based end-to-end network slicing, we design a new and standard-compliant system architecture with three functional layers as shown in Fig.~\ref{fig:network_architecture}. The network layer consists of network infrastructure devices, e.g., eNB and gNB, switches/routers, network functions, and edge and cloud servers.} The orchestration layer provides end-to-end resource management functions that allocate resources to network slices to ensure their performances according to SLAs. Domain managers are deployed in each technical domain, e.g., access, edge, transport, and core networks, with the action and state APIs to execute the orchestration. The domain managers realize the network slice subnet management functions (NSSMF) suggested by 3GPP~\cite{3gpp-nsmf}. The end-to-end orchestration and service assurance modules, which are similar to the 3GPP network slice management function (NSMF), aim to coordinate domain managers to ensure the end-to-end system performance of network slices~\cite{3gpp-nsmf}. \textcolor{black}{These modules are designed with interfaces that allow DRL agents to receive orchestration actions and feedback the slice performance statistics.} The intelligence layer implements DRL agents that learn the optimal resource orchestration policy for end-to-end network slicing. Since the machine learning components, i.e., DRL agents, are decoupled from orchestration functions, various machine learning techniques, e.g., safety learning, distributed learning and imitation learning, can be deployed without changing the underlying network slice functions in the orchestration layer. 

\section{DRL Challenges}
\label{sec:RL_challenge}
Although many existing works focus on applying DRL to solve various networking problems, deploying DRL-based network slicing solutions in practical end-to-end networks still faces multiple research challenges in terms of the performance assurance, solution scalability, and convergence speed of DRL.  

\subsection{Performance Assurance}
Network operators aim to satisfy service level agreements (SLAs) of slices with the minimum usage of cross-domain resources and thus reduce the operating expenses (OPEX). Realizing this goal is non-trivial because of the following research challenges.

\subsubsection{DRL explainability} 
To provide performance assurance, it is important to understand why a DRL-based slicing solution can lead to certain network performance. The difficulties come from two aspects. First, it is non-trivial to interpret the impact of the state space of DRL on the optimal policy. The state space should provide a comprehensive representation of network status to DRL agents. End-to-end networking systems generate a huge amount of measurement data from different domains. \textcolor{black}{There still has no clear path to follow and build the state space based on these network measurements in order to achieve the optimal network slicing policy. }

Second, the reason why the \textit{actions} generated by the DRL agent can achieve the optimal performance in long term is difficult to interpret. For a DRL policy, \textit{actions} are calculated through mathematical operations, e.g., addition, multiplex and activation in hidden layers and the output layer of a neural network. \textcolor{black}{Although the mathematical calculations are known, analyzing them can only provide very limited understandings about why an action is generated instead of the others, especially considering deep neural networks with heterogeneous layers such as convolutional and dropout. Without such interpretable knowledge, it's hard for network operators to directly control or modify \textit{actions} generated by DRL agents.}

\subsubsection{Exploration in policy optimization}
\textcolor{black}{DRL usually relies on random exploration mechanisms to find a better policy for improving long-term rewards. A practical useful exploration mechanism explores a nearby space of \textit{action} $\mathbf{a}_t$, e.g., $\mathbf{a}_t +\epsilon$ where $\epsilon$ is sampled from a normal distribution with zero mean and a predefined deviation}. As the new action deviates from the original action $\mathbf{a}_t$, the performances of network slices, e.g., throughput and delay, may be degraded to the extent that SLAs are violated. It is straightforward to limit the magnitude of the action exploration, e.g., $\mathbf{a}_t + clip(\epsilon, -H, H)$, where $H$ is a given maximum allowed deviation. However, this method reduces the exploration efficiency as the exploration is clipped. As a result, it may lead to a suboptimal policy or require more interactions learn the optimal policy in practice.

\subsubsection{Quantization error and parameterization}
DRL agents update their neural parameters by using gradient descent methods, e.g., SGD, Adam and Momentum~\cite{schulman2017proximal}, based on the collected historical trajectories (state-action-reward pairs).
Although the step size of gradient descent methods are usually small, e.g., 1E-3 or 1E-4, the performance of the DRL agent after one single policy update can be significantly different.
The reason is that DRL policies are often parameterized by millions of neurons, a slight change in each neuron can result in a dramatic discrepancy in output actions. Hence, when learning the optimal policy via interacting with a real end-to-end network, DRL agents may show changing performances during the learning phase. \textcolor{black}{As a result, it is challenging to assure a predictable performance of network slices using DRL-based slicing solutions.}

\subsection{Solution Scalability}

Learning-based networking mechanisms are designed to solve complex networking problems in a large scale. Hence, the scalability of DRL-based network slicing solutions determines whether they can be deployed in practical networking systems with heterogeneous infrastructures, e.g., base stations and servers, and dynamic service demands. The challenges of scaling DRL-based solutions up-and-down are from two aspects: communication and computing overhead, and neural network design.

\subsubsection{Communications and computing overhead}
DRL agents need to collect network measurements from distributed infrastructures to train the neural network~\cite{mao2019learning}. As the scale of the network increases, more measurement data need to be transmitted over the network, which may incur excessive communication overhead. 
According to OpenCelliD database, the number of base stations (e.g., GSM, UMTS, CDMA, LTE) of a medium city downtown (several square miles) could reach 1000.
Consider each base station transmits hundreds of parameters every millisecond, the transmission demands of such area will reach up to gigabits per second.
Meanwhile, managing a large-scale network also requires DRL agents to add more dimensions in their \textit{state} and \textit{action} spaces. As a result, the DRL agent consumes more computing resources and takes a longer time to learn the optimal policy. 

\subsubsection{Neural network design}
\textcolor{black}{DRL policies are parameterized by deep neural networks whose input and output dimensions are selected based on the \textit{states} and \textit{actions}~\cite{schulman2017proximal}, where the dimensions stay unchanged throughout the policy training. When the networking condition changes, e.g., new admitted network slices, both the \textit{states} and \textit{actions} space need to be revised to reflect this change.} As a result, the input and output dimensions of the neural network need to be updated, and the DRL policy has to be trained under the new networking condition. 
Although recently emerged recurrent neural networks (RNN) and graph neural networks (GNN) can handle the flexible sizes of inputs and outputs, their implementations in the context of DRL still need further investigations in terms of data efficiency, generalization and convergence.

\subsection{Convergence Speed of DRL}

DRL agents learn the optimal policy by directly interacting with real networking systems. To obtain the optimal policy, a DRL agent usually needs to be trained by a large number of online interactions. \textcolor{black}{Depending on how frequently the network measurements can be collected, it may take a very long time to achieve the convergence of the DRL policy}. The large-scale 5G network normally has a long orchestration interval~\cite{marquez2018should} as it involves the reconfiguration for various network devices and hardware.
Considering 15 minutes as a practical reconfiguration interval, the feedback of an orchestration action generated by the DRL policy can only be obtained every 15 minutes. Since a single policy update usually requires more than one thousand transitions~\cite{schulman2017proximal}, it needs, at least, a few days to achieve one policy update. 

\begin{figure*}[!t]
	\centering	
	\includegraphics[width=6.9in]{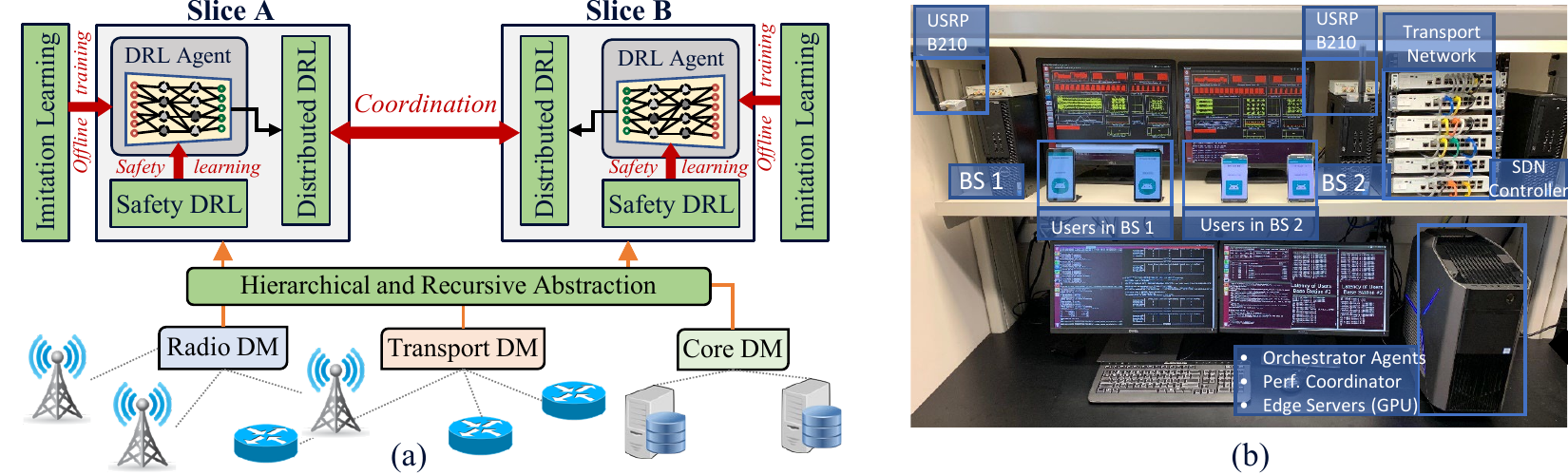}
	\caption{\small a) the explored techniques for addressing challenges; b) the network slicing testbed~\cite{liu2020edgeslice}.}
	\label{fig:solution_all}
\end{figure*}

\section{System Challenges}
\label{sec:system_challenge}
Deploying DRL-based network slicing solutions faces several system level challenges such as preparing networking data and providing appropriate network programmability for DRL.

\subsection{Networking Data for DRL}
DRL agents are trained with measurement data collected from heterogeneous and distributed network infrastructures. Preparing the data for DRL training can be a huge burden on networking systems.

\subsubsection{Data heterogeneity}
The openness evolution of 5G leads to the dis-aggregation of network components, which allows flexible network deployment strategies and increases the heterogeneity of these data in terms of formats, volume and time scales.
For example, 3GPP suggests splitting the gNB function into a central unit (CU) and a distributed unit (DU). The DU focuses on signal and data processing at PHY and MAC layers, and CU manages controlling functions at higher layers such as PDCP and RRC. O-RAN~\cite{o-ran} further split the gNB functions into CU, DU and radio unit (RU), where RU only provides functions related to radio frequency (RF) and Low-PHY processing such as fast Fourier transform (FFT), inverse FFT and physical random access channel (PRACH) extraction.
To efficiently manage measurement data from various network functions, O-RAN introduces the key performance measurement (KPM) function that defines a collection of information elements such as 5G QoS Identifier (5QI), QoS Class Identifier (QCI), time stamp, and Cell Object ID. These information elements contain highly heterogeneous data.
In addition, the service-based architecture of 5G core (5GC) implements new network functions such as access and mobility management function (AMF), session management function (SMF), and user plane function (UPF), which further increases the amount and heterogeneity of network data.

\subsubsection{Data processing capability}
Since a large amount of heterogeneous network data have to be collected and analyzed for DRL-based network slicing solutions, additional computing devices and functions are necessary to meet such data processing demands.  
\textcolor{black}{For example, the logging size of an operating gNB could easily exceed 1GB every minute~\cite{foukas2016flexran}, which is difficult to be transmitted from infrastructures to DRL agents.}
To reduce the data size of runtime data, the function of feature extraction, e.g., autoencoder, may be implemented to generate a concise representation.
The feature extractor usually achieves a better compression rate and accuracy of representation if more computing resources are enforced, e.g., denser neural network architectures.
As a result, there is a tradeoff between the deployed computation resources and the extraction performances, which requires further investigations.

\subsection{Network Programmability}

\textcolor{black}{To reconfigure network slices dynamically, network systems need to provide virtualization functions and programming interfaces to allow DRL agents to configure end-to-end network slices.}

\subsubsection{End-to-end infrastructure virtualization}
\textcolor{black}{The infrastructure virtualization is a key technology to provide necessary isolations among network slices, which assure that the performance and functions of a network slice are not affected by the operations of any other slices.}
The existing virtualization solutions are designed for individual technical domains, e.g., RAN, transportation networks, and edge and cloud computing.
For example, FlexRAN virtualizes physical resource blocks (PRBs) in the MAC layer as virtual radio bandwidth in RAN~\cite{foukas2016flexran}; OpenFlow allows the creation of virtual networks using software defined networking (SDN) in transport networks; virtual machine and docker container techniques enable the virtualization of computing resources in edge and cloud computing. 
It is, however, challenging to integrate virtualization solutions from different technical domains due to the heterogeneity of time scale, programming interfaces, and data models. 
\textcolor{black}{For instance, the PRBs in RAN can be modified in milliseconds, while the scaling of docker containers in edge computing requires seconds to take effect.}

\subsubsection{Network resource model}
The network resource model (NRM) is introduced to provide predefined interfaces based on either XML, JSON, or YANG to query and manage network resources~\cite{3gpp-nrm}. The NRM defines various network resources that allow efficient network management. For example, \textit{RRMPolicyRatio} defines \textit{rRMPolicyMaxRatio}, \textit{rRMPolicyMinRatio} and \textit{rRMPolicyDedicatedRatio}, which represent the maximum, minimum and dedicated resource usage quota for the associated \textit{rRMPolicyMemberList}.
A \textit{RRMPolicyMember} is defined by its pLMNId and sNSSAI (S-NSSAI).
Considering diverse performance requirements of services, e.g., reliability and delay, the NRM needs to support more configurable parameters, e.g., antenna array, transmit power, modulation and coding schemes.
However, how these parameters affect the performance of users and enable more customization features for slices are still open problems.

\section{DRL-based Network Slicing Solutions}
\label{sec:solution}

\textcolor{black}{In this section, we discuss several promising techniques, as summarized in Table I, to address the aforementioned algorithm and system challenges.}
Fig.~\ref{fig:solution_all} (a) illustrates how these technologies can be deployed for end-to-end network slicing. 
\begin{itemize}
    \item The safety DRL addresses the challenge of the performance assurance by tackling the unpredictable exploitation and random exploration, and assuring performance requirements of slices throughout the learning phase.
    \item \textcolor{black}{The distributed DRL addresses the scalability challenge by leveraging multi-agent DRL and allowing effective coordination among distributed agents.}
    \item The imitation learning addresses the challenge of DRL convergence by offline imitating a baseline policy and obtaining a good start point for online learning.
\end{itemize}

We implement and evaluate these solutions using the testbed shown in Fig.~\ref{fig:solution_all} (b).
The radio domain manager is designed based on OpenAirInterface (OAI) project with FlexRAN support~\cite{foukas2016flexran} and use Ettus USRP B210 as the RF front-end of a base station.
The transport domain manager is developed based on OpenDayLight (ODL) with OpenFlow 1.3 to control the transportation network composed of Ruckus ICX series SDN switches.
We use OpenAir-CN project to deploy core network and docker container technique to enable edge computing virtualization.
The DRL agents are developed by using PyTorch 1.5 with 3-layer fully-connected neural networks.

\subsection{Safety DRL}
The safety DRL aims to train the DRL agent to maximize the cumulative rewards while maintaining different constraints.
It is accomplished by integrating the Lagrangian primal dual method for statistical performance assurance and the baseline switching mechanism for avoiding instantaneous violations.

\textcolor{black}{The reward shaping method, which re-weights the reward if constraints are violated, has been widely used to penalize the DRL agent and guide its training.
However, it is difficult to choose the optimal weights especially under multiple constraints.
For example, too small weights leads to insufficient penalization for the violations of constraints, while too large weights would suppress the exploration of DRL.}
To address this problem, recent advances adaptively incorporate these constraints into the reward function by using the Lagrangian primal dual method~\cite{tessler2018reward}.
In particular, Lagrangian multipliers are introduced to re-weight the reward function, which are updated with the sub-gradient descent method in a larger time scale than that of resource orchestration.
On learning the resource orchestration policy, the reward function of the DRL agent is rewritten by the Lagrangian function, which integrates both the original objective and the weighted constraints by using the updated Lagrangian multipliers.
The constraints can be eventually satisfied by alternatively updating the multipliers and training the DRL agent.

\begin{figure*}[!t]
	\centering	
	\includegraphics[width=7.0in]{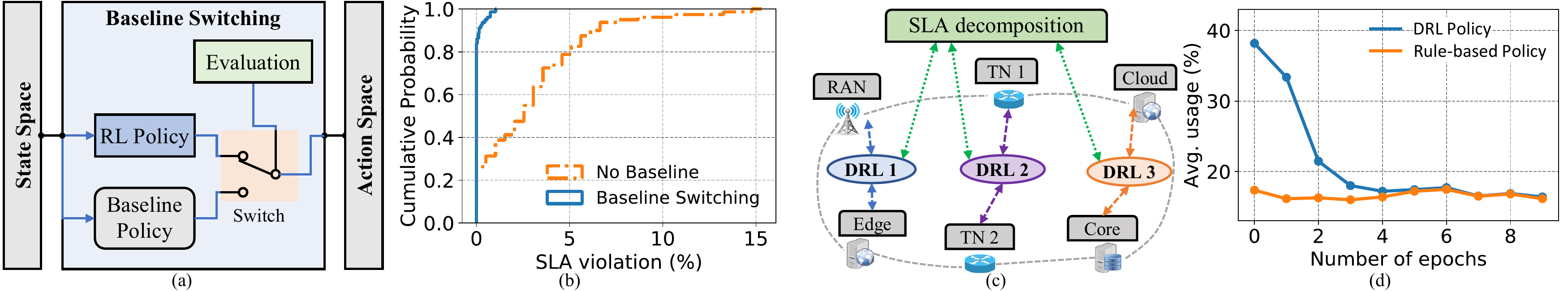}
	\caption{\small a) overview of baseline switching mechanism; b) SLA violation with baseline switching mechanism; c) illustration of distributed coordination; d) resource usage with imitation learning; }
	\label{fig:results}
\end{figure*}

The constraints, however, can still be violated before the Lagrangian primal dual method converges.
As shown in Fig.~\ref{fig:results} (a), a baseline switching mechanism is developed to enable the dynamic switching between the policy of the DRL agent and a baseline policy.
\textcolor{black}{We consider there is a baseline policy, e.g., rule-based or heuristic, which can guarantee the slice SLA but with higher resource usages~\cite{liu2021onslicing}.
The policy evaluation module is created to predict whether enforcing the current action generated by the DRL agent will break the slice SLA.
In particular, we design the module to learn the value of constraints under different states and actions.
This module can be implemented with DNN, where the input is the combination of state and action space, and the output is set to be the value of constraints, i.e., $c^t(\textbf{s}^t, \textbf{a}^t)$.}
If the prediction result is above zero with high confidence, the baseline policy will be invoked, which can maintain slice SLA at a cost of high resource usage.
Fig.~\ref{fig:results} (b) shows the cumulative probability of the SLA violation by two schemes, i.e., with and without the baseline switching mechanism.
\textcolor{black}{The SLA violation is defined if the constraints are violated, i.e., $c^t(\textbf{s}^t, \textbf{a}^t) \geq 0$. 
It can be seen that, without the baseline switching mechanism, the SLA violation can be as high as 15\%.
If the baseline switching mechanism is enabled, the SLA violation is reduced significantly ($\sim$1\%).}

\begin{table*}
\label{tb:summary}
\begin{center}
\small
\begin{tabular}{ | p{3.0cm}|p{5.5cm}| p{8cm} | } 
  \hline
  \textbf{Challenges}  &  \textbf{Existing Work}  &  \textbf{Our Solution} \\ 
  \hline
  Performance Assurance  &  unawareness of performance requirements, free action space exploration  &  \textbf{Safety DRL}: constraint-awareness update, baseline switching scheme  \\
  \hline
  Solution Scalability  &  communication \& computing overhead, fixed input/output DNN design  &  \textbf{Distributed DRL}: distributed multi-agents, coordination mechanism among agents\\ 
  \hline
  DRL Convergence  &  massive online interactions required, poor performance at early stage   &  \textbf{Imitation Learning}: offline imitate from baseline for online acceleration \\ 
  \hline
\end{tabular}
\captionof{table}{Summary of challenges and solutions}
\end{center}
\end{table*}

\subsection{Distributed DRL}
The distributed DRL aims to assure the scalability of the DRL agent when tackling dynamic networks, e.g., varying slice traffic and infrastructure deployment.
The multi-agent DRL is the common approach to scale the DRL agent in distributed networks, which creates multiple cooperative or competitive agents to achieve the global optima.
\textcolor{black}{For example, an individualized DRL agent can be created to orchestrate the resource for each network slice, where these agents are competitive due to the limitation of resource capacity.
This individualized slice agent scheme helps to resolve the dynamics of slice admission and departure, where the action space varies.}
The training of distributed agents can be accomplished by centrally aggregating their experiences and updating the policy of all agents simultaneously.

To orchestrate the cross-domain resources for end-to-end slices, multiple individualized DRL agents can be created in distributed infrastructures.
As illustrated in Fig.~\ref{fig:results} (c), the DRL agent 1 orchestrates RAN and edge networks, the DRL agent 2 controls transport networks, and the DRL agent 3 manages core networks and cloud computing resources.
These agents learn to allocate resources of infrastructures to slices independently, e.g., the DRL agent 2 allocates the bandwidth in transport networks.
As these agents only focus on their technical domains, the end-to-end performance requirement of slices may not be guaranteed if no collaborations.
\textcolor{black}{Thus, we propose an SLA decomposition module to decompose the end-to-end performance requirements of slices into different technical domains, which allow the DRL agents aware of local requirements.
For example, the end-to-end latency and reliability performance of a slice may be decomposed to local requirement in RAN, TN, CN, and Edge.
Then, the DRL agents in each domain is focused to allocate their resources to meet the local performance requirement of this slice.
In this scenario, the SLA decomposition module needs to balance the resource utilization in different domains, and is responsible for satisfying the end-to-end performance requirement. }

\subsection{Imitation Learning}
Imitation learning aims to train the DRL agent to mimic the behavior of a target agent, where the target agent can be either human, model-based rules or another DRL agent.
The main methods of imitation learning are behavior cloning, direct policy learning via interactive demonstrator, and inverse reinforcement learning.
Consider the network operator has a baseline policy, e.g., model-based rules, the imitation learning can train the DRL agent to mimic its behaviors, e.g., mapping the observed states to actions.
The advantage of offline imitation learning is that the transitions of the baseline policy are not expensive to obtain.
In Fig.~\ref{fig:results} (d), we offline train the DRL agent to imitate the behavior of the baseline policy by using the behavior cloning method.
Specifically, the offline training is accomplished by minimizing the action differences between the DRL policy and the baseline policy under different states.
\textcolor{black}{We see that the DRL agent uses a similar resource usage as the baseline policy does after several offline training epochs.
The DRL agent, after the imitation learning, serves as the start point for the online learning phase, which then continues learning and improving.}

\section{Future Directions}
\label{sec:future_directions}
We envision several promising techniques that may help to alleviate and resolve the aforementioned system and DRL challenges for end-to-end network slicing.

\subsection{Hierarchical and Recursive Abstraction}
The hierarchical and recursive abstraction (HRA) aims to adaptively abstract and manage resources in different domains both vertically and horizontally.
In the vertical direction, domain managers (DMs) in the same technical domain are hierarchically abstracted to provide high-level programmability.
For example, the coverage in RAN can be accomplished by abstracting radio resources in multiple base stations in the geographic area.
In the horizontal direction, DMs in different technical domains are recursively abstracted to provide end-to-end programmability.
For example, the end-to-end data rate can be achieved by abstracting virtual resources in both radio, transport and core DMs.
These DMs are responsible for enabling the network programmability of infrastructures, collecting the network state with standardized interfaces, and enforcing the management actions made by the DRL agents.

\subsection{Offline DRL} 
Offline DRL can be explored to offline train the DRL agents only based on collected online data sets. It helps to avoid expensive and unsafe online interactions with real networks. \textcolor{black}{The main difficulty is the distributional shift issue, i.e., although its function approximators (e.g., policy and value networks) is trained under one distribution (i.e., limited data set), it will actually perform on a different distribution (i.e., the real network).}

\subsection{Transfer Learning}
\textcolor{black}{
Transfer learning has shown a great potential to address challenges regarding scalability, model reproducibility, and sample efficiency~\cite{zhu2020transfer}.
The basic idea is to exploit prior learned knowledge to benefit the learning process in target tasks.
For example, several works have been done to transfer resource allocation policies from one network to another with similar settings, which accelerates the convergence speed of DRL agents~\cite{zhu2020transfer}.
Hence, the exploration of leveraging transfer learning in dealing with DRL challenges needs further research.
}
\subsection{Explainable AI}
The explainable artificial intelligence (XAI) aims to understand, interpret and explain the black-box DNN-based policies via different approaches, e.g., visual explanation, local explanations, illustrative examples, and simplifications. Various techniques, e.g., linear regression, decision trees, K-nearest neighbors, and Bayesian models, can be exploited to improve the explainability of DNN-based policies. For example, given a state observation, the Q-value of each possible action can be obtained from the Q network in the Deep Q-Network (DQN) agent. To maximize the long-term cumulative reward, the action with the highest Q-value is usually selected by the DQN agent. 

\section{Summary}
\label{sec:summary}
This article discusses DRL-based end-to-end network slicing. We have briefly studied the end-to-end resource orchestration problem and presented a new system architecture to enable DRL-based network slicing. We have also analyzed the challenges of deploying DRL-based solutions from the perspectives of both algorithm and system. \textcolor{black}{Moreover, we have explored and envision multiple promising techniques, e.g., safety and distributed DRL and transfer learning, for automating end-to-end network slicing.}

\bibliographystyle{IEEEtran}
\bibliography{ref/reference}

\begin{IEEEbiographynophoto}{Qiang Liu}
is an Assistant Professor in the School of Computing at the University of Nebraska-Lincoln. His research interests lie in the broad field of wireless communication, computer networking, edge computing, and machine learning.
\end{IEEEbiographynophoto}

\begin{IEEEbiographynophoto}{Nakjung Choi}
is the department head of mobile network systems and also DMTS (distinguished member of technical staff) in Nokia Bell Labs, Murray Hill, NJ, USA. His research focused on SDN/NFV/Cloud and 5G/IoT, especially end-to-end network and service automation.
\end{IEEEbiographynophoto}

\begin{IEEEbiographynophoto}{Tao Han}
(M’15-SM’20) is an Associate Professor in the Department of Electrical and Computer Engineering at the New Jersey Institute of Technology (NJIT) and an IEEE Senior Member. His research interests include mobile edge computing, machine learning, mobile X reality, 5G system, Internet of Things, and smart grid.
\end{IEEEbiographynophoto}

\end{document}